\documentclass[floats,aps,showpacs,twocolumn]{revtex4}

\usepackage{epsfig}
\usepackage{amsfonts}

-2.5cm   \def\8{\infty} \def\oh{\frac{1}{2}}
  
  \def\d{\partial}
 
\def\undertext#1{\vtop{\hbox{#1}\kern 1pt \hrule}}

\def\VEV#1{\left\langle\,#1\,\right\rangle}

\def\pbyp#1#2{\frac{\partial#1}{\partial#2}}

 \def\be{\begin{equation}}
  \def\ee{\end{equation}} \def\bea{\begin{eqnarray} & &}
  \def\eea{\end{eqnarray}}  \def\rf#1{(\ref{#1})}

\def\cH{{\cal H}}
\def\rf#1{(\ref{#1})}

\def\rfs#1{Eq.~(\ref{#1})}
\def\sign{{\rm sign}}

\begin{document}

\title{Magnon Localization in Mattis Glass}
\affiliation{test}
\author{Victor Gurarie$^{1,2}$ and Alexander Altland$^3$}
\affiliation{
$^1$Theoretical Physics, Oxford University, 1 Keble Road, Oxford
OX1 3NP, UK\\
$^2$Department of Physics, University of Colorado, CB 390, Boulder
CO 80309, USA \\ $^3$ Institut f\"ur Theoretische Physik,
Universit\"at zu K\"oln, Z\"ulpicher Str 77, K\"oln, Germany\\
}
\begin{abstract}
We study the spectral and transport properties of magnons in a
model of a disordered magnet called Mattis glass, at vanishing
average magnetization. We find that in two dimensional space, the
magnons are localized with the localization length which diverges
as a power of frequency at small frequencies. In three dimensional
space, the long wavelength magnons are delocalized. In the
delocalized regime in $3d$ (and also in $2d$ in a box whose size
is smaller than the relevant localization length scale) the
magnons move diffusively. The diffusion constant diverges at small
frequencies. However, the divergence is slow enough so that the
thermal conductivity of a Mattis glass is finite, and we evaluate
it in this paper. This situation can be contrasted with that of
phonons in structural glasses whose contribution to thermal
conductivity is known to diverge (when inelastic scattering is
neglected).

\end{abstract}


\maketitle

\section {Introduction}

For two reasons, the study of low energy excitations in spin
glasses -- spin glass {\it magnons} -- is met with notorious
difficulties: first, the energetic frustration characteristic for
glasses implies the existence of many nearly degenerate minima,
i.e. the conceptual status of ``small'' fluctuations forming on
top of any one of those configurations remains somewhat dubious.
Second, even if a sufficiently inert extremal configuration was
known, the solution of the appropriately linearized problem would
still pose a highly nontrivial problem.

To get the problem at least partially under control, magnons in
spin glasses are commonly described in the language of mean field
theory \cite{HS,AM}. Within this approach one finds that the
dispersion of the excitations forming on top of a background of
vanishing average magnetization
\begin{equation}
\label{m1} \omega(p) \propto |p|,
\end{equation}
is linear in analogy to the spinwave dispersion of
antiferromagnets.  Here, $\omega$ and $p$ denote frequency and
wave vector of the excitations, respectively. Mean field theory
further predicts that magnons in a glass with nonzero average
magnetization $m$ have a ferromagnet-like branch of excitations,
\begin{equation}
\label{m2} \omega(p) \propto p^2.
\end{equation}

However, in view of the difficulties alluded to above, there is no
reason {\sl a priori} why mean field theory qualifies to describe
spin glasses at all.  For example, it has been known for more than
a decade that mean field theory breaks down below the critical
dimension ${d}_c=2$ (see Refs.~\cite{Stinch,WHK,ACG}). (Yet above
the critical dimension arguments can be given which support the
mean field theory results Eqs.~\rf{m1} and \rf{m2} for the
excitation spectrum.)

Furthermore, an important question which mean field theory
completely fails to address is the transport properties of
magnons: while excitations in conventional magnetic materials
propagate ballistically, the disorder inherent to spin glasses
renders the dynamics of magnons diffusive and, eventually, leads
to mechanisms of localization. Clearly, these phenomena cannot be
described in terms of a spatially uniform mean field \cite{fn1}.

It is the main objective of the present paper to introduce an
alternative approach to the problem which is not burdened by these
limitations. Developed in close analogy to the field theory
approach to electron dynamics in disordered solids, the formalism
below can be employed to address both spectral {\it and}
localization properties of magnons. On the other hand, it has
nothing to say about the first problem mentioned above,
identification and analysis of reference ground states. For this
reason, we chose to introduce the approach on a prototypical
variant of a spin-glass, the so-called Mattis glass, for which
this problem simply does not exist.

To prepare the definition of the Mattis glass, let us recall that spin
glasses~\cite{MPV} are usually described by the exchange Hamiltonian
\begin{equation}
\label{hm}  H = -\oh \sum_{ij \alpha} J_{ij}  S_i^\alpha
S_j^\alpha,
\end{equation}
where $i$, $j$ refer to nearest neighbor sites on some lattice,
$J_{ij}$ are random exchange constants, and $\hat S^\alpha,
\alpha=1,2,3$ is a vector representing the spin. (In this paper we
assume that $|{\bf S}|\equiv S\gg 1$ is sufficiently large so that the
spin system can be treated as classical.) The spin equations of motion
are then given by,
\begin{eqnarray}
  \label{eqm}
&&  \pbyp{S}{t} = [H,S],\nonumber\\
&&\hspace{.5cm} \left[S^\alpha_j, S^\beta_k\right]
= i \delta_{jk} \sum_{\gamma} \epsilon^{\alpha \beta \gamma}
S^\gamma_k,
\end{eqnarray}
where $\epsilon^{\alpha \beta \gamma}$ is the usual antisymmetric
tensor.

As mentioned above, one of the major difficulties hampering
analytical progress on the problem posed by (\ref{eqm}) is that
the ground state(s) of the Hamiltonian are not known. For this
reason, we consider here a simplified variant of (\ref{hm}), the
so-called Mattis glass. The Mattis glass is defined by,
\begin{equation}
\label{Mattis}
 H = - {J \over S} \sum_{jk \alpha} \xi_j \xi_k S^\alpha_j
S^\alpha_k,
\end{equation}
i.e. a spin glass Hamiltonian for which the exchange constants $J_{jk}
\sim \xi_j \xi_k$ {\it factorize}.  Here, $J$ is a constant parameter
setting the energy units, $j,k$ are nearest neighbor sites of a
$d$-dimensional lattice, and $\xi_j$ are bimodally distributed random
variables taking value $1$ with probability $p$ and $-1$ with
probability $1-p$. This factorization entails that a degenerate family
of spin configurations minimizing the Hamiltonian can readily be
written down:
\begin{equation}
\label{ground} S_i^{\alpha (0)} = S~{\rm sign}\left(\xi_i \right)
n^\alpha,
\end{equation}
where $n$ is a unit vector with arbitrary direction (due to
rotational invariance of \rfs{Mattis}). Thus the Mattis glass is a
model with a disordered ground state but without the frustration
inherent to usual spin systems.  To understand the physical
features of Mattis glass magnons we  need to explore the equations
of motion (\ref{eqm}), linearized around (\ref{ground}).  (Under
the presumed condition $S\gg 1$ anharmonic fluctuations, i.e.
magnon {\it interactions} can safely be neglected.) We emphasize
that this part of the analysis is of similar complexity as the
study of magnons in ``real" spin glasses. Indeed, arguments can be
given to the effect that  magnons in real and Mattis glasses,
resp., behave by and large equivalently (see Ref.~\cite{GA}).

The first to study magnon propagation in the Mattis glass was
David Sherrington. In a series of papers \cite{Sher} he found the
spectral density of magnons in the Mattis glass and their
scattering length. Additionally, Ref.~\cite{Stinch} is an
important work studying magnons in a 1D Mattis glass. It has to be
mentioned that any random one dimensional spin chain with nearest
neighbor interaction is automatically a Mattis glass, and since in
1D the scattering length coincides with the localization length
all the transport properties of Mattis magnons in 1D are known
\cite{Sher,Stinch}. Because of that, in this paper we will
concentrate on dimensionalities higher than one. Before turning to
the methodological aspects of the analysis, let us summarize our
main results and relate them to earlier work on similar problems.

{\it Summary of Results:} In this paper we will characterize the
behaviour of magnons in terms of their dispersion relation $\omega(p)$, the
frequency dependent
localization length, $l(\omega)$, and the so-called density of frequencies
\begin{equation}
\label{eq:DoS}
\rho(\omega) = {1 \over N} \sum_{n=1}^N
\delta\left(\omega-\omega_n \right),
\end{equation}
where $\omega_n$ are the frequencies of the magnon modes and $N$ is
the total number of modes in the glass. Borrowing terminology from the
physics of disordered electron systems, we will henceforth refer to
$\rho(\omega)$ as the density of states (DoS). In a way to be
formulated precisely below, the localization length, $l(\omega)$, is a
measure for the exponential decay of magnon modes at frequency
$\omega$.
As measurable observables related to these
quantities we will consider the thermal conductance of the glass, and
its specific heat.

We first note (see the discussion in the end of the next section)
that if the average magnetization of a Mattis glass is nonzero,
the behavior of its magnons at very low frequencies closely
resembles that of phonons in a {\it structural} glass. Since
phonons in structural glasses have already been discussed in
Ref.~\cite{SJS}, in this paper we concentrate mostly on the Mattis
glass with zero average magnetization, or with $p=1/2$. In some
instances below where we do consider $p \not = 1/2$, this will be
explicitly stated.

For the dispersion relation of the  un-magnetized Mattis
glass we find
\begin{eqnarray}
  \label{dispersion}
  \begin{array}{lll}
\displaystyle  {\rm Re}~\omega \propto p,& {\rm Im}~\omega \propto
p^{2} & {\rm in \ }3d,\cr \displaystyle {\rm Re}~\omega \propto {p
\over \sqrt{\log\left({\Lambda \over p}\right)}}, & {\rm
Im}~\omega \propto {p \over\log\left({\Lambda \over p}\right)^{3
\over 2}}, & {\rm in \ }2d.
  \end{array}
\end{eqnarray}
This is consistent with the discussion in Refs.~\cite{WHK} and
\cite{ACG} where it is argued that ${\rm Im}~ \omega \propto
p^{d-1}$ above the critical dimension $d>d_c=2$. This behavior can
be contrasted with the behavior of phonons in structural glasses
whose dispersion relation is
\begin{equation}
\label{dispersion2} {\rm Re}~\omega \propto p, \qquad {\rm
Im}~\omega \propto  p^{{d}+1}.
\end{equation}
Eq.~\rf{dispersion2} is often interpreted as a manifestation of
Raleigh scattering. A similar formula also holds for glasses with
nonzero average magnetization $M$ (but with the substitution
$\sqrt{\omega}$ instead of $\omega$).

In Ref.~\cite{Sher} self consistent diagrammatic methods  were
applied to derive the DoS
\begin{eqnarray}
  \label{eq:DoSM}
  \begin{array}{ll}
\displaystyle \rho(\omega) \propto \omega \left| {\log\left(\omega
\right)} \right| & {\rm in \ }2d, \cr \displaystyle \rho(\omega)
\propto \omega^2 & {\rm in \ } 3d.
  \end{array}
\end{eqnarray}
Our field theoretical analysis below will confirm this result.
Also notice that Eqs.~(\ref{eq:DoSM}) coincide with the DoS
deduced from mean field theory \rfs{m1} (up to a logarithmic
prefactor in 2d).

The central result of this paper regards the localization
properties of magnons. We find that in 2d magnons are localized on
the frequency dependent scale
\begin{equation}
l(\omega) \propto \omega^{-{1 \over 16 \pi}}.
\end{equation}
(In contrast, phonons in structural glasses or, equivalently,
magnons in Mattis glass with nonzero $M$, are subject to a weaker
localization mechanism leading to an exponentially diverging
localization length \cite{SJS}.)  At length scales below the
localization length the magnons move diffusively.  In the 3d case
we find that the dynamics is diffusive no matter how small the
frequency, i.e. there is no localization.

In both $2d$ and $3d$, the diffusion  constant diverges as the
frequency of magnons goes to zero:
\begin{equation}
D(\omega) \propto {1 \over \omega^{{d}-1}}.
\end{equation}
(Compare this with $D(\omega) \propto \omega^{-{d}-1}$ for
structural glasses \cite{SJS}).  Dispersive observables such as
the thermal conductivity are related to the product of diffusion
constant and density of states, $\rho(\omega) D(\omega)$ (weighted
by the specific heat, $C(\omega)$, and integrated over frequency.)
A glance at Eq. (\ref{eq:DoSM}) shows that this quantity is an
integrable function at small $\omega$. As an important
consequence, we find that, independent of dimensionality, the
thermal conductivity of magnons in a Mattis glass is finite.
Specifically, for $2d$,
\begin{equation}
\label{eq:thermal1} \kappa={k^2 T \over 12  \hbar} \log \left( {4
\pi \Lambda^2 J^2 \over \alpha  k^2 T^2 } \right).
\end{equation}
Here, $\Lambda$ is a momentum cutoff, reciprocal to the minimal
wavelength of magnons, and $\alpha$ measures the correlation
volume of spins (often $\alpha \propto \Lambda^{-{d}}$ for short
range correlated spins). Notice that the conductivity is only
weakly disorder dependent. This is a consequence of the fact that
the diffusion constant/DoS scale linearly/inversely linear with
the disorder strength. (Similar behavior is observed, e.g., for
quasiparticles in high temperature superconductors~\cite{ASZ}.)

In $3d$, the thermal conductivity is given by the fully disorder
independent result,
\begin{equation}
\label{therm3} \kappa = {\Lambda \over 9 \pi} {k^2 T \over \hbar}.
\end{equation}
Owing to the absence of localization, the thermal {\it
conductance} is obtained from (\ref{therm3}) by a trivial
multiplication by the system size: Ohm's law.

In this paper we also discuss the Mattis glass with nonzero
magnetization. The magnon's low frequency behavior is closely
related to that of phonons in a glass. Thus the density of states
is $\rho(\omega) \propto \omega^{{{d}-2}\over 2}$, in agreement
with \rfs{m2}. The thermal conductivity is infinite because of the
weak scattering at small frequencies, as can be seen from the
frequency dependent diffusion constant $D(\omega) \propto
1/\omega^{{d} \over 2}$. Finally, at frequencies higher than a
certain frequency $\omega_c$, the magnons' behavior crosses over
to that of a glass with vanishing magnetization. In $2d$,
$\omega_c=J M/\alpha$, where $J$ is the spin-spin interaction
constant multiplied by the square of the lattice constant and
$\alpha$ is the correlation volume of spins, see below. In $3d$,
$\omega_c=J M/(\alpha \Lambda)$, where $\Lambda$ is the inverse
lattice spacing.

This concludes our preliminary summary of results. The rest of the
paper is organized as follows. In the next section we derive the
magnon equations of motion for Mattis glass. We then formulate the
field theory approach (section \ref{sec:field_theory}) and use it
discuss magnon localization and transport properties in $2d$ and $3d$
(section \ref{sec:results}.)

\section{The Equations of Motion}

The most straightforward way to introduce magnons on a technical level
is by Holstein-Primakoff transformation \cite{Sher}. For those spins
$S_j$ with $\xi_j>0$, we write, in the harmonic approximation (site
indices suppressed for notational transparency),
\begin{equation}
S^z=S-a^\dagger a, \ S^+\approx a \sqrt{2 S}, \ S^- \approx
a^\dagger \sqrt{2 S},
\end{equation}
while for those whose $\xi_j<0$ we write
\begin{equation}
S^z=-S+a^\dagger a, \ S^+\approx a^\dagger \sqrt{2 S}, \ S^-
\approx a \sqrt{2S},
\end{equation}
where $a$, $a^\dagger$ are Holstein-Primakoff bosons.  Substituting in
\rfs{Mattis} and expanding to second order in $a$, $a^\dagger$, one
obtains the quadratic Hamiltonian \cite{Sher}
\begin{eqnarray}
\label{energy}
&&\delta H = {J\over 2} \sum_{jk=1}^N \left( \matrix{
a^\dagger_j & a_j } \right) \left( \matrix {h_{jk} & \Gamma_{jk}
\cr \Gamma_{jk} & h_{jk}} \right) \left( \matrix{ a_k \cr
a^\dagger_k } \right)\equiv\nonumber\\
&&\hspace{.5cm}\equiv {J\over 2} \psi^\dagger {\cal H} \psi,
\end{eqnarray}
where
\begin{equation}
\cH =\left( \matrix {h & \Gamma \cr \Gamma & h} \right), \ \psi =
\left( \matrix{ a \cr a^\dagger } \right),
\end{equation}
and $h$ and $\Gamma$ are $N \times N$ symmetric matrices defined in
the following way
\begin{eqnarray}
  \label{eq:1}
&& h_{jj}=2d, \ h_{j,j+\mu}=-\oh \left( 1+\xi_j
  \xi_{j+\mu} \right), \\
&& \Gamma_{j, j+\mu} = \oh \left(1 - \xi_j
  \xi_{j+\mu} \right). \nonumber
\end{eqnarray}
Here $j+\mu$ refer to the nearest neighbor sites of the site $j$.  As
discussed in Ref.~\cite{GC}, the magnon equation of motion then takes
the form
\begin{equation}
\label{eqm1} \cH \psi = \omega \left( \matrix{ \openone_N & 0 \cr 0 &
-\openone_N} \right) \psi,
\end{equation}
where $\openone_N$ is the $N$-dimensional identity matrix.  To
proceed, it is convenient to perform a unitary rotation
\begin{eqnarray}
&&\cH - \omega \left( \matrix{ \openone_N & 0 \cr 0 & -\openone_N } \right)
\rightarrow\nonumber\\
&&\hspace{2cm}\to U^\dagger \left( \cH - \omega \left( \matrix { \openone_N & 0
\cr 0 & -\openone_N} \right) \right) U,\nonumber\\
&&\psi\to U^\dagger \psi
\end{eqnarray}
where the unitary matrix $U$ is defined as
\begin{equation}
U = \oh \left( \matrix { \Xi-\openone_N & -\Xi-\openone_N \cr \Xi+\openone_N & \openone_N -\Xi
} \right),
\end{equation}
and  $\Xi_{jk} = \delta_{jk} \xi_k$. As a result, \rfs{eqm1}
assumes the block-diagonal form
\begin{equation}
{J \over 2} \left( \matrix { h-\Gamma & 0 \cr 0 & h-\Gamma }
\right) \psi = \omega \left( \matrix{ -\Xi & 0 \cr 0 & \Xi }
\right) \psi,
\end{equation}
where the disorder-independent combination $h-\Gamma$ is but the
$d$-dimensional lattice Laplacian $\Delta_{jk}$. (As a side remark
we mention that the unitary equivalence ${\cal H}\sim -\Delta>0$
states the positivity of the operator ${\cal H}$ --  for a bosonic
problem,
 a necessary stability condition.)  Thus we find
\begin{equation}
\label{eom:discrete}
-{J \over 2} \sum_k \Delta_{jk} \psi_k = \pm \omega \xi_j \psi_j.
\end{equation}
Here the two signs refer to the upper/lower half of the vector
$\psi$. Physically, they correspond to two magnon branches of
opposite frequency.  In the context of Mattis glass
Eq.~(\ref{eom:discrete}) first appeared in Ref.~\cite{Stinch}. As
a last step, we take the continuum limit of this equation. Trading
the discrete index $j$ for a continuous coordinate $x$, we arrive
at \cite{Junits}
\begin{equation}
-{J \over 2} \Delta \psi = \pm \omega \xi(x) \psi,
\end{equation}
where $\Delta$ is the continuum Laplacian and the discrete random
variable $\xi_j$ got replaced by a function $\xi(x)$, randomly taking
values $+1$ or $-1$ with probabilities $p$ and $1-p$ at different
points in space.

As it is hard to deal with bimodally distributed
variables, we are going to replace $\xi(x)$ according to $\xi
\rightarrow M + V(x)$, where $M$ is the average magnetization per spin
$M = 2p - 1$ and $V(x)$ is a random Gaussian variable with zero mean
and correlation
\begin{equation}
\label{Vcorr} \VEV{V(x) V(y)} = \alpha~\delta(x-y).
\end{equation}
Remembering the definition of $V$ as a continuous version of the
variables $\xi_k$, we deduce that the parameter $\alpha$ has the
meaning of the correlation volume of spins in the Mattis glass.
(For uncorrelated spins, $\alpha$ is of the order of the
elementary lattice volume.)

In view of universality of extended random systems with respect to
changes in the microscopic realization of the disorder, we trust that
taking a continuum limit and modelling the randomness according to
(\ref{Vcorr}) are permissible simplifications. At any rate, the final
form of the equations of motion on which our further analysis will be
based reads as
\begin{equation}
\label{eqm2} \left(  {J \over 2} \Delta + \omega  M + \omega V(x)
\right) \psi = 0,
\end{equation}
where we have suppressed the $\pm$ sign in front of the frequency.

Structurally, \rfs{eqm2} bears resemblance to the Schr\"odinger
equation of a quantum  particle in a random
potential. The crucial difference is, however, that what would have
been an energy in that equation is now  proportional to
magnetization, and what would have been the disorder strength is
proportional to the frequency of magnons.

\rfs{eqm2} is also very similar to the main equation of
Ref.~\cite{SJS} where phonons in structural glasses were studied.
However, what is $M+V(x)$ in \rfs{eqm2} was the mass density $m(x)$ in
Ref.~\cite{SJS}, and as such, it was a strictly positive quantity. In
the Mattis glass, on the other hand, $M+V(x)$ (being a continuum
analog of the bimodal distribution $\xi_j \in \{-1,1\}$) can and must
be negative for some values of $x$. Yet the methods employed in
Ref.~\cite{SJS} required to take $M+V(x)$ as a random Gaussian
variable centered around $M\equiv \langle m(x)\rangle$. Without
discussing whether that would invalidate any of the conclusions of
Ref.~\cite{SJS} with respect to the structural glasses, we immediately
deduce that the properties of magnons in Mattis glasses with nonzero
average magnetization $M$ have effectively been already derived, and
Ref.~\cite{SJS} can be consulted to find out how that was done. (For
completeness, the main results had been summarized in the previous
section.) In what follows we concentrate on the complementary case of
the Mattis glass with vanishing magnetization $M=0$ (or $p=1/2$), with
the exception of section IV, where $M \not =0$ will also be
considered.

\section{Field theory}
\label{sec:field_theory}

\subsection{Green functions}
All relevant information about the eigenvalue problem defined by
(\ref{eqm1}) and (\ref{eqm2}) is contained in the advanced and
retarded Green functions
\begin{equation}
\label{eq:gfp}
G^{\pm} ={1 \over \omega + {J \over 2} V^{-1} \Delta \pm i
\epsilon},
\end{equation}
where $\epsilon>0$ is infinitesimal and we have set the average
magnetization to zero, $M=0$. An equivalent (see Appendix), but
for our purposes more convenient representation of the Green
function reads as
\begin{equation}
\label{eq:gf} G^{\pm} = \left[ \omega V  +{J \over 2} \Delta \pm i
\epsilon~{\rm sign\,} \omega \right]^{-1} V.
\end{equation}

The two quantities we shall focus on in the following are
\begin{eqnarray}
  \label{eq:cfs}
&&  C_1(\omega) \equiv \langle G^+(\omega;{\bf x},{\bf x}) \rangle,\\
&&  C_2(\omega,\Omega;|{\bf x}-{\bf y}|) \equiv \nonumber\\
&&\hspace{1cm}\equiv \left\langle G^+(\omega+\Omega/2;{\bf x},{\bf
y})\, G^-(\omega-\Omega/2;{\bf y},{\bf x}) \right\rangle,
\nonumber
\end{eqnarray}
where the averaging $\langle \dots \rangle$ is over ``disorder" $V$.
As usual, the average DoS can be represented in terms of the Green
functions as
\begin{equation}
\label{eq:DoS1} \rho(\omega) = -{1\over \pi} {\, \rm Im\,}
C_1(\omega).
\end{equation}
$C_2(\omega, {\bf p})$
can be used to calculate diffusion constant of magnons. At small
$\Omega \ll \omega$ and small ${\bf p}$,
\begin{equation}
\label{eq:diffu} C_2(\omega, \Omega; {\bf p}) \approx {4 \pi
\rho(\omega) \over D(\omega) {\bf p}^2 + i \Omega },
\end{equation}
where $D(\omega)$ is the diffusion constant of magnons at
frequency $\omega$.

The thermal conductivity of the system is then given by \cite{SJS}
\begin{equation}
\label{eq:thermal} \kappa = \int_0^\infty d\omega~\rho(\omega)
C(\omega, T) D(\omega),
\end{equation} where $C$ is the specific heat of magnons at
frequency $\omega$,
\begin{equation}
C(\omega, T) = {(\hbar \omega)^2 \over k T^2 \left( 2 \sinh {\hbar
\omega \over 2 k T }\right)^2 }.
\end{equation}

To compute the disorder average involved in the definition of the
correlation functions $C_{1,2}$, we employ the field theoretical
formalism of the nonlinear $\sigma$-model, an approach that has been
met with tremendous success in the field of disordered fermion
physics. In fact, the application of this field theory to ``glassy"
problems is by no means original. In a pioneering work, John,
Sompolinsky and Stephen~\cite{SJS} attacked the related problem of
phonon localization in structural glasses by the same formalism.
However, while at first sight the presence (phonons) or absence
(Mattis glass magnons) of the parameter $M$ in (\ref{eqm2}) may appear
to be of minor significance, the opposite is the case. In fact,
essential elements of the structure of the theory depend on this
point, which is why very different physical results are obtained (cf.
discussion above.)

To prepare the averaging over the disorder, we consider the
 supersymmetric generating functional:
\begin{equation}
\label{gen1} {\cal Z}[V] = \int D \Psi \exp(-S[\bar \Psi,\Psi]),
\end{equation}
where the action $S$ is defined as
\begin{equation}
  \label{eq:psiS}
  S=i \bar \Psi \left[ {J \over 2}
\Delta + \left( \omega + {\Omega \over 2} \sigma_3^{AR} \right) V
+ i \epsilon \sigma_3^{AR} \right] \Psi.
\end{equation}
Here, $\Psi\equiv \{\Psi^\alpha\}$, $\alpha=1,\dots,8$ is an
eight-component vector field, half of whose components are complex
variables, while the remaining components are anticommuting (Grassmann
variables). Of the commuting (anticommuting) variables, one half
refers to the retarded sector of the theory, the other half to the
advanced sector. The simultaneous presence of both sectors is
indicated by the presence of the Pauli matrix $\sigma_3^{\rm AR}$
operating in advanced-retarded space. While our so far counting
accounts for four different types of variables
(commuting-anticommuting/advanced-retarded), a further doubling of the
number of integration variables is required~\cite{Efetov} by the time
reversal invariance of the problem (technically, the fact that we are
dealing with a symmetric operator.)

Referring for a comprehensive introduction to the apparatus of
supersymmetry in statistical physics to Ref.~\cite{Efetov} we here
merely mention that the rational behind introducing a
supersymmetric structure is that ${\cal Z}[V]=1$ is automatically
unit-normalized. (The operator determinants resulting from the
integration over commuting/anticommuting variables, resp., cancel
each other.) Not having to worry about determinantal prefactors,
the average over the Gaussian distribution of the disorder becomes
a straightforward operation. Again referring for a detailed
discussion to Ref.~\cite{Efetov}, we here just note that a
Hubbard-Stratonovich decoupling of the four-fermion 'interaction'
generated by the disorder averaging results in
\begin{eqnarray}
\label{action}
&&{\cal Z} \equiv \langle {\cal Z}[V]\rangle
= \int {\cal D} R \exp (-S[R]),
\end{eqnarray}
where $R=\{R^{\alpha\beta}\}$ is an eight-dimensional matrix field,
whose action reads as
\begin{equation}
  \label{Raction}
   S[R]= {1 \over 4 \alpha \omega^2}\int {\rm str\,}(R^2)
 + {1 \over 2} {\rm str\,ln\,}
{\cal G}[R]^{-1}.
\end{equation}
Here, str is the so-called supertrace~\cite{Efetov}, and
\begin{equation}
{\cal G}[R]^{-1} = {J \over 2} \Delta  +  \left(1+ {\Omega \over 2
\omega} \sigma_3^{AR} \right) R.
\end{equation}

Eq.~(\ref{action}) represents an (admittedly complicated)
representation of unity, ${\cal Z}=1$.  To make use of the
formalism, we need to relate our correlation functions $C_{1,2}$
to the generating functional.  Indeed, it is straightforward to
verify that the one point correlation function $C_1$ is obtained
by computing the expression
$$
C_1(\omega)= i \left\langle \left \langle S^{1} ({\bf x}) S^{1\ast}({\bf x})
\right \rangle_{\psi} V({\bf x})\right\rangle_V,
$$
where $\langle \dots \rangle_\psi$ denotes the functional average
over the Gaussian action (\ref{eq:psiS}) and $S^{1}$ refers to the
first component of the commuting sector of the field $\Psi$. After
the disorder average this is equivalent to computing
\begin{equation}
\label{RC1} C_1(\omega)=\alpha \omega \VEV{S^1({\bf x})
S^{1*}({\bf x})
 \bar \Psi({\bf x}) \Psi({\bf x} )}.
\end{equation}

As for the 'two-point' function $C_2$, certain care has to be
exercised with the sign of the frequency arguments: phenomena like
diffusion and localization are observed in the limit of small
frequency {\it difference} $|\Omega|\ll \omega$.  Assuming that the
offset frequency $\omega>0$ is positive, the analytic structure (cf.
Eq. (\ref{eq:gf})) of the Green function is thus determined as
indicated in (\ref{eq:cfs}). Accordingly, the two-point function can
be represented as
\begin{eqnarray}
  \label{RC2}
&& C_2(\omega, \Omega; |{\bf x}-{\bf y}|)= \alpha^2 \left\langle
S^1({\bf x}) S^{1*} ({\bf y})
  S^2({\bf y}) S^{2*} ({\bf x}) \right. \times \nonumber \\
&&\hspace{1cm}
  \times \bar \Psi({\bf x})\left(\omega+{\Omega \over 2} \sigma_3^{AR} \right)
\Psi({\bf x} ) \times \\
 &&\hspace{1cm} \times \left. \bar \Psi({\bf
y})\left(\omega+{\Omega \over 2} \sigma_3^{AR} \right) \Psi({\bf
y} )
   \right \rangle. \nonumber
\end{eqnarray}
Here $S^1$ and $S^2$ stand for advanced and retarded components of
the commuting sectors of the vector $\Psi$.

To actually evaluate the functional expectation values (\ref{RC1})
and (\ref{RC2}), we need to reduce the (exact) reformulation of
the problem, Eqs.~(\ref{action}) and (\ref{Raction}), down to a
better manageable effective low energy field theory. This
reduction is achieved by a gradient expansion around the spatially
uniform stationary phase configurations of the action
(\ref{Raction}). These reference states are determined by solution
of the equation
\begin{eqnarray}
  \label{mfeq}
&&   {\delta S[\bar R]\big|_{\Omega=0}\over \delta \bar R({\bf x})}=0
\Leftrightarrow\\
&&\hspace{1cm}\Leftrightarrow  \bar R =  \alpha \omega^2
\left({1\over 2\pi}\right)^{{d}}\int  {d^{d} p\over {J \over 2}
p^2 - \bar R},\nonumber
\end{eqnarray}
where in the second equality $p$ is momentum.  (Owing to $\Omega\ll
\omega$, the dependence of the mean field configurations on the
frequency mismatch $\Omega$ is weak, which is why  $\Omega$ is
neglected in (\ref{mfeq}).)

The simplest solutions to Eq.~(\ref{mfeq}) acquire the matrix-diagonal
form
$$
\bar R = q_1 + i q_2 \sigma_3^{\rm AR},
$$
where $q_{1,2}$ are real parameters. The detailed form of
$q_{1,2}$ depends on the dimensionality of the problem.  To avoid
reiterations, we therefore first outline the general (independent
of dimensionality) architecture of the theory before discussing
the cases ${d}=1,2,3$ separately.

All we will rely upon in the following is that, at sufficiently
small $\omega$,  $q_1$ is parametrically larger than $q_2$. The
significance of this relation becomes clear when we note that,
physically, $\bar R$ plays the role of a self-consistent Born
(SCBA) self-energy of the problem (in diagrammatic language, the
self energy calculated neglecting diagrams with crossing impurity
scattering lines.)

Indeed, (\ref{mfeq}) is structurally equivalent to the familiar form
of the SCBA self energy equation.  The statement $q_1 \gg q_2$ thus
implies that the energy {\it shift} acquired by impurity induced
virtual transitions from one magnon mode into others exceeds the
energy {\it broadening}, i.e. the instability of magnons in a
non-translationally invariant environment.  Again alluding to the
formal analogy to the Green functions of disordered electrons, $q_1
\sim E_F$ plays the role of the ``Fermi energy" of the problem, and
$1/q_2\sim \tau$ is the analog of the inverse fermionic scattering
time, $\tau$. The fact that $q_1/q_2 \sim E_F \tau \gg 1$ implies that
we are working with the analog of a weakly disordered system. At the
same time, the parameter $q_1/q_2 \gg 1$ stabilizes the construction
of the low energy field theory to be outlined momentarily.

\subsection{Construction of a low energy action}
\label{sec:constr-low-energy}

The fact that in the limit $\Omega\to 0$ the action (\ref{Raction})
is isotropic in the internal matrix space of the theory implies that
$\bar R$ is but a single element of an entire manifold of
solutions of the saddle point equation. Indeed, any configuration
$$
T \bar R T^{-1} \equiv q_1 + i q_2 T \sigma_3^{\rm AR} T^{-1}
$$
represents another solution. Here
$T\in {\rm OSp\,}(4|4)$ is an eight-dimensional supermatrix lying in
the supergroup ${\rm OSp\,}(4|4)$, i.e. the maximal group manifold
compatible with the internal symmetries of the problem\cite{Efetov}.
Generalizing to slowly fluctuating rotations, we obtain the matrix
field
\begin{eqnarray}
  \label{Qdef}
&&  Q({\bf x})\equiv T({\bf x})\sigma_3^{\rm AR} T^{-1}({\bf x})
\in\nonumber\\
&&\hspace{1cm}\in
  {\rm OSp\,}(4|4)/{\rm
  OSp\,}(2|2) \times {\rm OSp\,}(2|2)
\end{eqnarray}
as the central degree of freedom of the theory.

The final step in the construction is to expand the action in slow
fluctuations of the field $Q({\bf x})$. The finite cost of these
fluctuations is due to (a) their spatial variation and (b) the
presence of the so far neglected frequency mismatch, $\Omega$.
Fortunately, the job of actually determining the resulting
contributions to the action has been done before \cite{Efetov}.
All we need to do is carefully identify the parameters of the
present problem with those of its electronic analog; the algebraic
structure of the two problems is almost identical. Indeed,
substituting the 'soft' configurations $R=q_1 + i q_2 Q$ into the
action (\ref{Raction}), we obtain
\begin{equation}
  \label{SQs1}
 S[Q]={\,\rm str\,ln\,}\left[{J \over 2} \Delta  +q_1 + iq_2 Q + {\Omega
 \over 2 \omega}
 \sigma_3^{AR}
 \left(
 q_1  + i q_2 Q  \right) \right].
\end{equation}
It is useful to relate this expression to the action of the
fermionic problem (see Ref.~\cite{Efetov})
\begin{equation}
  \label{SQf}
 S^{\rm f}[Q]={\,\rm str\,ln\,}\left({1\over 2m}\Delta  +E_F + {i\over
 2\tau} Q + \Omega \sigma_3^{\rm AR}\right).
\end{equation}
Comparison of the two actions leads to the list of identifications
summarized in the table I below.

\begin{table*}
\label{table}
\begin{tabular}{|l|l|l|}\hline
quantity & fermions & magnons\\\hline
Fermi energy & $E_F$  &$q_1$\\
scattering time  & $\tau$ & $(2q_2)^{-1}$\\
mass         & $m$      & $J^{-1}$\\
\hline\hline diffusion constant, $D_0$ & $D_0={2 E_F \tau\over
{d}m}$& $D_0={ J q_1\over {d} q_2
}$\\
density of states, $\nu$ & $\nu = {\Gamma^{({d})}   \over
(2\pi)^{d}} (2m E_F)^{{d}-2\over 2} {d} m$& $ \nu =
{\Gamma^{({d})} \over (2\pi)^{d}} \left({2 q_1\over
J}\right)^{{d}-2\over 2} {{d} \over J}$\\\hline 
\end{tabular}
\caption{Comparison of basic (upper part) and derived (lower part)
scales of the problem. In the fifth row, $\nu$ denotes the DoS per
volume and $\Gamma^{(d)}$ is the volume of the ${d}$-dimensional
unit sphere.}
\end{table*}

Outgoing from (\ref{SQf}), an {\it effective} low energy action
\begin{eqnarray}
  \label{Seff}
  S^{\rm f}_{\rm eff}[Q]= {\pi \nu\over 8} \int d^{d}r\, {\rm
  str}\left(D_0
  \partial Q \partial Q + 2i \Omega Q \sigma_3^{\rm AR}\right).
\end{eqnarray}
can be derived by leading order expansion in the parameters
$\Omega\tau\ll 1$ and $l/L\ll 1$, where $L$ is representative of
the length scales we wish to probe. Eq. (\ref{Seff}) contains the
bare values of DoS, $\nu$, and diffusion constant, $D_0$, of the
fermionic problem. These quantities are related to the parameters
of the prototypical action, Eq. (\ref{SQf}) as also summarized in
the table I. Details of the derivation can be found in
Ref.~\cite{Efetov}.

Quite analogously, starting from the action \rfs{SQs1} the
effective low energy action of the magnon problem can be obtained
as
\begin{eqnarray}
  \label{Seff1}
&&  S_{\rm eff}[Q]=  \int  d^{d}r\, {\rm str}\left(
  {\pi \Gamma^{({d})} \over (2 \pi)^{d}} {q_1^{{d}
  \over 2} \over 8 q_2 } \left({2 \over J}\right)^{{d}-2 \over 2}
  \partial Q \partial Q +  \right. \nonumber\\
&&\hspace{2cm}+  \left. i{q_1 q_2 \over \alpha \omega^3} \Omega Q \sigma_3^{\rm
AR}\right).
\end{eqnarray}
We next proceed to discuss what can be learned from this
representation of the problem.  The first quantity we would like to
calculate is the magnon density of states $\rho(\omega)$. This can be
extracted from the correlation function $C_1$ with the help of
\rfs{eq:DoS1} using the relations \rfs{RC1} and \rfs{mfeq}. The result
is
\begin{equation}
\label{eq:fullDoS}
\rho(\omega) = {4 q_1 q_2 \over \pi \alpha
\omega^3}.
\end{equation}
Using Eq. (\ref{eq:fullDoS}), we rewrite \rfs{Seff1} in a
form similar to the effective action of the fermion problem,
Eq. (\ref{Seff}):
\begin{equation}
S_{{\rm eff}}= {\pi \rho(\omega) \over 8} \int d^{d} r \left[
D(\omega) \d Q \d Q + 2 i \Omega Q \sigma_3^{AR} \right],
\end{equation}
Here $D(\omega)$ is the diffusion constant of magnons
at frequency $\omega$,
\begin{equation}
D(\omega) = {\pi \Gamma^{({d})} \over (2 \pi)^{d}} \left({2 q_1
\over J}\right)^{d-2 \over 2}{\alpha \omega^3 \over 4 q_2^2}.
\end{equation}
Finally, the product $\rho(\omega) D(\omega)$, relevant for the
calculation of the thermal conductivity (cf. Eq. \ref{eq:thermal})
reads as
\begin{equation}
\rho(\omega) D(\omega) =
  { \Gamma^{({d})} \over (2 \pi)^{d}} {q_1^{{d}
  \over 2} \over  q_2 } \left({2 \over J}\right)^{{d}-2 \over 2}.
\end{equation}
Calculating the correlation function $C_2$ using \rfs{RC2}, we
indeed reproduce the expression \rfs{eq:diffu}.

Finally, we note that the dispersion relation of magnons is given
by the poles of the Green's function $G(\omega, {\bf p})$ and can
be extracted from
\begin{equation}
\label{eq:magdisp1} {J \over 2} p^2 = q_1+i q_2.
\end{equation}

Since the functional form of the mean field parameters $q_{1,2}$
depend on the dimensionality of space, we next consider the cases
${d}=1,2,3$ separately.

\section{Results}

\label{sec:results}
\subsection{$1\,{d}$ case}
\label{sec:1-rm}

Strictly one-dimensional Mattis glasses are amenable to transfer
matrix techniques and have been discussed before us \cite{Stinch}.
Nevertheless we would like to try to analyze them here using the
techniques proposed in this paper. Our conclusion, however, is
going to be that, as always for strictly one dimensional problems,
the sigma model description breaks down.

In ${d}=1$, the momentum integral appearing in the saddle point
equation (\ref{mfeq}) can readily be done and we obtain
\begin{equation}
\label{mfD1} \bar R = \left({\alpha^2 \omega^4 \over
2J}\right)^{1/3}\cos(\pi/3)\left(1+ i \tan(\pi/3) \sigma_3^{\rm
AR}\right).
\end{equation}
Substitution of this result into (\ref{eq:fullDoS})  leads to an
SCBA DoS
\begin{equation}
  \label{rho1}
  \rho(\omega) = -{1\over \pi}{\,\rm Im\,}(C_1(\omega)) \sim \omega^{-1/3},
\end{equation}
divergent at small frequencies. Eq.~(\ref{rho1}) in fact coincides
with the exact expression for the density of states obtained in
Ref.~\cite{Stinch} by the transfer matrix formalism. A direct
application of SCBA to the 1D Mattis glass was attempted even
earlier, in Ref.~\cite{Sher}. However, from a purist point of
view, the result does not seem quite trustworthy. A glance at
(\ref{mfD1}) shows that real ($q_1$) and imaginary ($q_2$) part of
the SCBA self energy are actually of the same order. On the other
hand, the SCBA as such is stabilized by the smallness of the
parameter $q_2/q_1$. Put differently, in a one-dimensional
environment the SCBA, and for that matter the construction of our
low energy field theory, are ill-founded. For this reason, we are
not going to discuss the case ${d}=1$ any further and turn to
${d}=2$ instead.

\subsection{ $2{\,d}$ case}

In ${d}=2$, the saddle point equation (\ref{mfeq}) takes the form
\begin{equation}
\bar R = { \alpha  \omega^2 \over (2 \pi)^2} \int {d^2 p \over {J
\over 2} p^2 -\bar R}
\end{equation}
Cutting the logarithmic divergence of this integral by introducing a
momentum cutoff $\Lambda$, we obtain
$$
\bar R = {\alpha  \omega^2\over 2\pi J} \log \left(-{\Lambda^2 J
\over 2 \bar R}\right).
$$
To logarithmic accuracy, this equation is solved by $ \bar R = q_1
+ i q_2 \sigma_3^{\rm AR}$, where
\begin{eqnarray*}
&&  q_1 = {\alpha \omega^2 \over 2\pi J} \log\left({\pi \Lambda^2
J^2
  \over \alpha \omega^2}\right) \\
&& q_2 = {\alpha \omega^2 \over 2 J}.
\end{eqnarray*}
Notice that we are now on safe ground inasmuch as
\begin{equation}
\label{eq:cond} q_1/q_2 \sim \log\left({ \pi \Lambda^2 J^2 \over
\alpha \omega^2} \right)\gg 1,
\end{equation}
or in other words, as long as the frequency is small enough,
\begin{equation}
\omega \ll {\Lambda J \over \sqrt{\alpha}}.
\end{equation}
Armed with $q_1$ and $q_2$ we can now calculate the DoS, the
diffusion constant and the thermal conductivity. With the help of
\rfs{eq:fullDoS} we find
\begin{equation}
\label{eq:DoS2D} \rho(\omega) = {\alpha \omega \over  \pi^2 J^2}
\log\left[ { \pi \Lambda^2 J^2 \over \alpha \omega^2} \right].
\end{equation}
This indeed coincides with the density of states derived in
Ref.~\cite{Sher}. Looking at the dispersion relation
\rfs{eq:magdisp1} with the help of $q_2 \ll q_1$ we see that
\begin{eqnarray*}
&&{\rm Re} \,\omega =  \sqrt{4 \pi \over \alpha \log \left(
{\Lambda^2
\over p^2} \right)} {J \over 2} p, \\
&&{\rm Im} \, \omega = \left({\pi \over  \alpha^{1 \over
3}\log\left( {\Lambda^2 \over p^2} \right) }\right)^{3 \over 2} {J
\over 2}p.
\end{eqnarray*}
Next, the diffusion constant is given by
\begin{equation}
\label{eq:dif2D} D(\omega)={J^2 \over 4 \alpha \omega}.
\end{equation}
This allows us to calculate the thermal conductivity by using
Eqs.~\rf{eq:DoS2D}, \rf{eq:dif2D}, and \rf{eq:thermal}. Quite
remarkably, the integral in \rfs{eq:thermal} is convergent at
small frequency and we obtain a closed formula for the magnon
thermal conductivity in $2d$
\begin{equation}
\label{eq:therm2D} \kappa={k^2 T \over 12  \hbar} \log \left( {
\Lambda^2 J^2 \over \alpha k^2 T^2 } \right) .
\end{equation}
As a consequence of \rfs{eq:cond}, this formula works if the
logarithm in it is large, or $T \ll \Lambda J/k \sqrt{\alpha}$.

Finally, we would like to calculate the localization length of
phonons at frequency $\omega$. To do that, we evaluate the
$\Omega=0$ effective field theory action \rfs{Seff1} to find
\begin{equation}
S_{\rm eff} = {1\over 32 \pi} \log \left( { \pi J^2 \Lambda^2
\over \alpha \omega^2} \right) \int d^{d} r ~{\rm str\,}
\partial Q
\partial Q.
\end{equation}
In two dimensions, the supersymmetric $\sigma$-model is
known\cite{Efetov} to flow to a disordered phase. The magnon
localization length is the length scale at which the corresponding RG
group equations renormalize the coupling constant of the effective
field theory,
\begin{equation}
 {1 \over 32 \pi } \log \left[ { \pi \Lambda^2 J^2 \over \alpha
\omega^2} \right]
\end{equation}
down to values of the order of 1. That gives (compare with
Ref.~\cite{SJS})
\begin{equation}
l(\omega) \propto \exp \left\{ {1 \over 32 \pi } \log \left[ {
\pi \Lambda^2 J^2 \over \alpha \omega^2} \right] \right\}
\end{equation}
or
\begin{equation}
l(\omega) \propto \omega^{-{1 \over 16 \pi}}.
\end{equation}
The localization length is divergent as a power law of the
frequency. This is in contrast to the behavior of phonons in
structural glasses \cite{SJS} and, by extension, of magnons in
Mattis glass with nonzero overall magnetization. There the
localization length diverges much faster with decreasing
frequency, as an exponential of the inverse frequency square. In
other words, in these systems, static disorder is less efficient a
scattering agent than in the Mattis glass with vanishing
magnetization.

\subsection{$3d$ case}

Conceptually, the analysis of the $3d$ cause parallels the
discussion of the previous section. We therefore restrict
ourselves to a brief statement of the key formulae.

In analogy to the $2d$ case, we solve the  mean field equations
\rfs{mfeq} to
\begin{eqnarray}
\label{eq:q12}
&&q_1 = {\alpha \omega^2 \over  \pi^2 J} \Lambda,\nonumber\\
&&q_2= \sqrt{ \alpha^3 \omega^6 \Lambda \over 2 \pi^4 J^4}.
\end{eqnarray}
At sufficiently small frequencies $\omega$
\begin{equation}
\label{eq:cond2} q_1/q_2 \propto {J \sqrt{\Lambda} \over \omega
\sqrt{\alpha}} \gg 1,
\end{equation}
and the effective field theory approach works in this case as
well. The DoS follows as
\begin{equation}
\label{eq:3ddos} \rho(\omega) = 8 {\left( \alpha \Lambda
\right)^{3 \over 2} \over 2^{3 \over 2} \pi^5 J^3} \omega^2,
\end{equation}
in agreement with the general mean field theory result above the
critical dimension  ${d}
> {d}_c=2$. The dispersion relation of magnons is
\begin{eqnarray*}
&& {\rm Re} \, \omega = \sqrt{ \pi^2 J^2 \over 2 \alpha \Lambda} p
\\
&& {\rm Im} \, \omega = {\pi \over 4 \Lambda} \sqrt{ \pi^2 J^2
\over 2 \alpha \Lambda} p^2,
\end{eqnarray*}
in agreement with the general result ${\rm Re} \, \omega \propto
p$, ${\rm Im} \, \omega \propto p^{{d} -1}$. The diffusion
constant is given by
\begin{equation}
\label{eq:3ddif} D(\omega) = { \pi^2 \over 6 \sqrt{2}} {J^3 \over
\sqrt{\Lambda} \alpha^{3 \over 2} \omega^2}.
\end{equation}
The thermal conductivity can be found with the help of
\rfs{eq:thermal} to give
\begin{equation}
\label{eq:therm3D} \kappa={\Lambda \over 9 \pi} {k^2 T \over
\hbar} .
\end{equation}
In $3d$, the sigma model at weak bare coupling $E_F\tau \sim
q_1/q_2\gg 1$ is known to flow to a metallic phase where the
conductance $K \sim \kappa L$ is Ohmic. Accordingly, the heat
conductance of a sample of linear size $L$ will be given by
\begin{equation}
K={\Lambda L \over 9 \pi} {k^2 T \over \hbar}.
\end{equation}
Due to the condition, \rfs{eq:cond2}, the temperature has to satisfy
\begin{equation}
T \ll {J \over k} \sqrt{\Lambda \over \alpha}.
\end{equation}

\subsection{Quasi-$1\,{d}$ case}
\label{sec:quasi-1-rm}

Quasi $1{d}$ systems are highly anisotropic such that the
extension in one direction (the ``longitudinal" direction) is far
in excess of the ``transverse" extensions. Here we will consider
such a three-dimensional ``wire'' made of Mattis glass.

Let us denote the width of the wire, when measured in units of
lattice spacings, $N$. Then the actual width of the wire is given
by $N/\Lambda$, where $\Lambda$, as before, is the inverse lattice
spacing. $N=1$ corresponds to the strictly one-dimensional
problem, while here we will consider the $N\gg 1$ case.

As a result of having finite $N$, the transverse momentum is
quantized in units of (roughly) $\Lambda/N$. The saddle point
equation \rfs{mfeq} must now involve both integration over
longitudinal momentum and the summation over the transverse ones.
Fortunately it is possible to go back to three dimensional
integration if
\begin{equation}
q_2 \gg J \left( {\Lambda \over N}\right)^2.
\end{equation}
Once the integration is three dimensional, we can borrow $q_2$
from \rfs{eq:q12} to find
\begin{equation}
J \sqrt{\Lambda \over \alpha} \ll \omega N^{2 \over 3}.
\end{equation}
This is the condition which the frequency and the number of
channels must satisfy for the 3$d$ density of states be applicable
to the quasi-1$d$ geometry.

At the same time, for the formalism to work the frequency cannot
be larger than $J \sqrt{\Lambda/\alpha}$, due to \rfs{eq:cond2}.
We see that as a result, the frequency must lie in the interval
\begin{equation}
\label{eq:freqint} {1 \over N^{2 \over 3}} \ll {\omega \over J}
\sqrt{\alpha \over \Lambda} \ll 1,
\end{equation}
which is always possible at large enough $N$.

Once the magnon frequency lies in the interval given by
\rfs{eq:freqint}, the magnons are described by the 3$d$ sigma
model derived in the previous subsection. However, for
sufficiently narrow wires and long times the diffusion of magnons
becomes purely one-dimensional. For that to happen, we need to
take the frequency $\Omega$ to be much less then the inverse
Thouless time $D(\omega)/(N/\Lambda)^2$. In other words, using
\rfs{eq:3ddif} we find
\begin{equation}
\label{eq:Thouless} \Omega \ll {\Lambda^{3 \over 2} J^3 \over N^2
\alpha^{3 \over 2} \omega^2}.
\end{equation}

Our conclusion is that in the long quasi-1$d$ wires the magnons
whose frequencies satisfy \rfs{eq:freqint} diffuse and localize
purely one-dimensionally for times longer than the inverse
$\Omega$ given in \rfs{eq:Thouless} above, even though their
density of states and their diffusion constant as a function of
frequency $\omega$ are given by the 3$d$ expressions
Eqs.~\rf{eq:3ddos} and \rf{eq:3ddif}.

\section{Nonvanishing Magnetization}

\subsection{Phonon-Magnon crossover}

It is also instructive to see how the magnons behave when the
average magnetization is nonvanishing $M>0$. While the properties
of very low frequency magnons coincide with those of phonons in
structural glasses discussed in \cite{SJS}, at somewhat higher
frequencies the behavior crosses over to the one indistinguishable
from that of the zero magnetization magnons. It is this phenomenon
that we would like to study in this section.

The starting point of the analysis is Eq.~\rf{eqm2} with $M>0$. It is
fairly straightforward to repeat the analysis of section
\ref{sec:field_theory} of the paper to find the following effective
action, the analog of \rfs{SQs1},
\begin{eqnarray} &&
  \label{SQsnonz}
 S[Q]={\,\rm str\,ln\,}\left[{J \over 2} \Delta  +M\omega +q_1 + iq_2 Q + \right.
 \\ &&
 \left. +{\Omega
 \over 2 \omega}
 \sigma_3^{AR}
 \left(
 M \omega + q_1  + i q_2 Q  \right) \right]. \nonumber
\end{eqnarray}
Accordingly, the self-consistent Born mean field equation \rfs{mfeq} changes
to
\begin{equation}
\bar R =  \alpha \omega^2 \left({1\over 2\pi}\right)^{{d}}\int
{d^{d} p\over {J \over 2} p^2 - \bar R-M \omega }.
\end{equation}
The solution of this equation has, as before, the form
\begin{equation}
\bar R=q_1 + q_2 \sigma^{AR}.
\end{equation}
Finally, the structural form of Green functions and observables
themselves changes. Specifically, the Green's functions now have the form
\begin{equation}
G^{\pm}=\left[ \omega \left(V  +M \right) +{J \over 2} \Delta \pm
i \epsilon~{\rm sign\,} \omega \right]^{-1} \left(V+M \right).
\end{equation}
Comparison with the fermionic effective action \rfs{SQf} shows that
the role of the Fermi energy is now played by $M \omega +q_1$. Two
distinct regimes can now be identified: for $M\ll
q_1(\omega)/\omega$, the magnon properties are indistinguishable from
those of Mattis glass magnons in the absence of a magnetized
background. In contrast, for $M \gg q_1(\omega)/\omega$, the magnons
resemble phonons in structural glasses.  As we shall see in a moment,
at sufficiently low frequencies, the second scenario is effectively
realized.

To demonstrate how this works, let us solve the SCBA equation
\rfs{SQsnonz} at ${d} =2$. We find
\begin{eqnarray}
&&q_1={\alpha \omega^2 \over  2 \pi J} \log \left( {\Lambda^2 J
\over 2 q_1 +  2 M \omega } \right),\nonumber\\
&&q_2={\alpha \omega^2 \over 2 J}.
\end{eqnarray}
If
\begin{equation}
{\alpha \omega^2 \over 2 \pi J} \log \left( {\Lambda^2 J \over 2 M
\omega} \right) \ll  M \omega,
\end{equation}
the magnons are in the phonon regime. This will happen if
\begin{equation}
\omega \ll { J M \over \alpha}.
\end{equation}
Conversely, at $\omega \gg J M/\alpha$, the magnons are
effectively in the $M=0$ regime studied above in this paper.
However, the condition $q_1 \gg q_2$ must also apply in order for
the formalism developed in this paper to work. This is equivalent
to $\omega \ll \Lambda J/\sqrt{\alpha}$.  The consistency
condition is thus $\Lambda J / \sqrt{\alpha} \gg J M / \alpha$ or
\begin{equation}
\label{eq:condit} M \ll \sqrt{\Lambda^2 \alpha}.
\end{equation}
The average magnetization $M\le 1$, while $\Lambda^2 \alpha > 1$.
We see that \rfs{eq:condit} always applies.

Analogously, at ${d}=3$, we find that the magnetization $M$ can be
neglected if
\begin{equation}
\omega \gg { M J \over \alpha \Lambda}.
\end{equation}
Its consistency with $q_1 \gg q_2$ gives $M \ll \sqrt{\Lambda^3
\alpha}$.

\subsection{The Phonon regime}

At small enough frequency, the magnons behave in a way reminiscent
of phonons in structural glasses. Even though most of the results
for those were obtained in Ref.~\cite{SJS} we would like to
present some of them here, in part for completeness, and in part
since there are still some distinctions between phonons and
magnons. This mostly stems from the fact that the basic equation
for phonons in glasses, while practically identical with
\rfs{eqm2}, had $\omega^2$ in place of $\omega$.

The main feature simplifying the magnetized theory at small frequency
is that $M$ always dominates the effective ``Fermi" energy. As a
result, it is possible to neglect $q_1$ in all the calculations. The
Green's functions can be approximately calculated as
\begin{equation}
G^{\pm} \approx M \left[ \omega \left(V  +M \right) +{J \over 2}
\Delta \pm i \epsilon~{\rm sign\,} \omega \right]^{-1} .
\end{equation}
Because of this structure, the problem enjoys a complete analogy to
the problem of disordered fermions, and the results of the treatment
of the latter can be directly translated into the language of the
former. This fact has already been successfully exploited in
Ref.~\cite{SJS}.

In particular, it is clear that the DoS will
coincide with what we called $\nu$ in the previous sections of the
paper,
\begin{eqnarray}
&& \rho(\omega) =  \nu(\omega) = {1 \over (2 \pi)^{d}} \int d^{d}
p~ \delta \left( {J \over 2} p^2 - M \omega \right) =\cr && =
{\Gamma^{({d})} \over (2 \pi )^{d}} {{d} \over J} \left( { 2 M
\omega \over J} \right)^{{d} -2\over 2} .
\end{eqnarray}
This coincides with the spectrum of ferrimagnets \cite{HS,ACG}.

Concentrating on the case of two dimensions ${d}=2$, we find that
as before $q_2=\alpha \omega^2 /(2 J)$, which leads to the
diffusion constant
\begin{equation}
D(\omega)\equiv D_0(\omega) = {2 J^2 M \over \alpha \omega}
\end{equation}
The integral $\int d\omega~\rho(\omega) D(\omega)$ is
logarithmically divergent, leading to the infinite thermal
conductivity of magnons at nonzero magnetization. A similar
phenomenon is also observed with phonons in structural glasses
(although the degree of divergence is different there). The
physical consequence of this latter phenomenon is that in real
magnets with nonvanishing magnetization the thermal conductivity
must be dominated by inelastic scattering processes (completely
neglected in this paper) rather than by elastic collisions.

We finally note that  the $2d$ dispersion relation reads ${\rm Im}
\,\omega \propto p^3$, in agreement with the general formula
\rfs{dispersion2}. Similarly, in $3d$ $q_2\propto \omega^{5 \over
2}$ resulting in ${\rm Im} \,\omega \propto p^4$, again in
agreement with \rfs{dispersion2}. The diffusion constant goes as
$D\equiv D_0(\omega) \propto 1/\omega^{3/2}$, leading to the same
divergent behavior of $D(\omega) \rho(\omega) \propto 1/\omega$ as
in $2d$.


\section{Conclusions}

In this paper we studied the spectral and transport properties of
magnons in a Mattis glass with vanishing average magnetization. We
find that in $3d$ their motion is diffusive and that results in
finite thermal conductivity given by \rfs{eq:therm3D}. In $2d$ the
thermal conductance is also finite and given by \rfs{eq:therm2D}.
This could be contrasted with the behavior of phonons in
structural glasses (and with magnons in a Mattis glass with
nonzero magnetization), whose contribution to thermal conductance
is infinite (when phonon-phonon scattering is neglected).

Of course, one may object that the Mattis glass as such represents a
somewhat artificial model system. We believe, however, that in spite
of its artificial definition, this system captures much of the physics
of magnon propagation in generic disordered magnets.  Arguments to the
effect that the results derived in this paper may be of wider
significance will be presented in a forthcoming publication\cite{GA}.

\section{Acknowledgement}

The authors are grateful to J.T. Chalker, M. Zirnbauer and A.
Andreev for many stimulating discussions. V.G. was supported by
the EPSRC Advanced Research Fellowship. A.A. acknowledges support
by Sonderforschungsbereich SFB/TR 12 of the Deutsche
Forschungsgemeinschaft.

\section{Appendix}

We would like to show that
\begin{equation}
\label{eq:theorem}  \left[ \omega V + {J \over 2} \Delta \pm i
\epsilon V \right]^{-1}= \left[ \omega V + {J \over 2} \Delta \pm
i \epsilon ~\sign \, \omega \right]^{-1}
\end{equation}
in the limit when $\epsilon$ is small. Once this is shown, the
expression for the Green's function \rfs{eq:gf} follows
immediately.

The proof is based on the fact that $-{J\over 2} \Delta$ is a
positive definite operator. For simplicity, let us first consider
an analog of \rfs{eq:theorem} written for real numbers as opposed
to operators. Take
\begin{equation}
\label{eq:epsilon} {1 \over \omega v - d \pm i v \epsilon},
\end{equation}
where $v$ is a real number and $d$ is a positive real number. Take
$\omega>0$. Then if $v$ is negative, $\omega v-d$ is also
negative, and $\epsilon$ can be completely neglected. If $v$ is
positive, the product $v \epsilon$ can be replaced by $\epsilon$,
since we are interested in the limit $\epsilon \rightarrow 0$
anyway. In other words, regardless of the sign of $v$,
\begin{equation}
\label{eq:equa} {1 \over \omega v - d \pm i v \epsilon} = {1 \over
\omega v - d \pm i  \epsilon}, \ \omega>0
\end{equation}
as $\epsilon$ is taken to zero. For negative $\omega$, the logic
can be repeated to result in
\begin{equation}
\label{eq:equa1} {1 \over \omega v - d \pm i v \epsilon} = {1
\over \omega v  - d \mp i  \epsilon}, \ \omega<0.
\end{equation}
These can be combined to give
\begin{equation}
\label{eq:equa2} {1 \over \omega v - d \pm i v \epsilon} = {1
\over \omega v - d \pm i  \epsilon~\sign \, \omega},
\end{equation}
at arbitrary sign of $\omega$.

The operator generalization of \rfs{eq:equa2} can be derived by similar
reasoning: take $\omega>0$ as in the text above. To show that
\rfs{eq:theorem} holds, we calculate matrix elements of
\begin{equation}
G^{-1}\equiv \omega V+{J\over 2} \Delta \pm i \epsilon V
\end{equation} in the basis where $\omega V+{J \over 2} \Delta$ is
diagonal. Denoting the eigenvalues of $\omega V+{J\over 2} \Delta$
as $\lambda_a$ and the eigenvalues of ${J\over 2} \Delta$ as
$-\mu_a$ ($\mu_a>0$), we find
\begin{equation}
\label{eq:matr} \left( \omega V + {J \over 2} \Delta \pm i
\epsilon V \right)_{ab} = \lambda_a \delta_{ab} \pm {i \epsilon
\over \omega} \left( \lambda_a \delta_{ab} + U_{ac} \mu_c
U^\dagger_{cb} \right),
\end{equation}
where the matrix elements of ${J\over 2} \Delta$ in the reference basis are
written as $-U_{ac} \mu_c U^\dagger_{cb}$, $U$ is some unitary
matrix and summation over the index $c$ is implied. In what follows,
it is crucial that $U_{ac} \mu_c U^\dagger_{ca}>0$, or in other words,
that the diagonal entries of a positive matrix have to be positive.

The eigenvalues $\lambda_a$ are generally nonzero. As we tune the
parameter $\omega$, the eigenvalues $\lambda_a$ can go through zero
one at a time.  If all of $\lambda_a$ are nonzero, $\epsilon$ can be
taken to zero directly in \rfs{eq:matr}, and we obtain
\begin{equation}
G_{ab} = {\delta_{ab} \over \lambda_a}.
\end{equation}
If one of $\lambda_a$ is zero, more care is needed. The
off-diagonal matrix elements of $G$ can be found as the ratio of
the appropriate minors of $G^{-1}_{ab}$ to the determinant of
$G^{-1}_{ab}$. It is easy to see that the determinant of
$G^{-1}_{ab}$ is of the order of $\epsilon$, while the minors are
at least of the order of $\epsilon$ as well (or they could be of
higher order in $\epsilon$). As a result, the off-diagonal matrix
elements of $G_{ab}$ are either constant as $\epsilon$ goes to
zero, or vanishing. However, being a constant for a particular
value of $\omega$ (for which one of the eigenvalues $\lambda_a$
vanished) and zero at all other values of $\omega$ is tantamount
to being zero (as in $\lim_{\epsilon \rightarrow 0} {i\epsilon
/(\lambda + i \epsilon)} = 0$ as a function of $\lambda$).

As far as the diagonal matrix elements of $G$ are concerned, they
are also given by the ratio of the minors to the determinant of
$G^{-1}$. The main contribution to both in the limit $\epsilon
\rightarrow 0$ is the product of the appropriate diagonal elements
of $G^{-1}$. It is easy to see that in that limit the matrix
elements of $G_{aa}$ are equal to $1/\lambda_a$ except when
$\lambda_a=0$. For that one vanishing eigenvalue,
$$
G_{aa} = {\omega \over i \epsilon \left(U_{ac} \mu_c
U^\dagger_{ca} \right)}.
$$
Since $U_{ac} \mu_c U^\dagger_{ca} >0$, and $\omega>0$, and since
$\epsilon$ is taken to zero, we can replace this by
$$
G_{aa} ={1 \over i\epsilon}
$$
in that limit. So indeed, the coefficient in front of $\epsilon$
can simply be put to 1 from the very beginning, as in
\rfs{eq:theorem} for $\omega>0$.

Repeating the same argument for $\omega<0$, we arrive at
\rfs{eq:theorem} for arbitrary sign of $\omega$.


\begin{thebibliography}{99}

\bibitem{HS} B.I. Halperin, W.M. Saslow, Phys. Rev. B {\bf 16},
2154 (1977)

\bibitem{AM} A.F. Andreev,
Sov. Phys. JETP {\bf 47}, 411 (1978)

\bibitem{Stinch} R.B. Stinchcombe, I.R. Pimentel, Phys. Rev. B
{\bf 38}, 4980 (1988)

\bibitem{ACG} J.T. Chalker, V. Gurarie, cond-mat/0305445, to be
published in Phys. Rev. B

\bibitem{WHK} C.C. Wan, A.B. Harris, D. Kumar, Phys. Rev. B{\bf 48}, 1036
(1993)


\bibitem{fn1}  Mean field theory becomes exact for spin glasses with
infinite range interactions. In that case, one finds \cite{BM}
that the frequency dependent density of magnon modes (for a
precise definition of this quantity, see
Eq.\protect~(\ref{eq:DoS}) below) scales as $\rho(\omega) \sim
\omega^{3/2}$. However, this powerlaw is at variance with the
behavior observed for ``real" spin glasses.

\bibitem{BM} A.J. Bray, M.A. Moore, J. Phys. C: Solid State Phys.,
{\bf 14}, 2629 (1981)

\bibitem{MPV} M. Mezard, G. Parisi, M. Virasoro,
Spin Glass Theory and Beyond, World Scientific Lecture Notes in
Physics, Vol 9


\bibitem{GA} V. Gurarie, A. Altland, in preparation

\bibitem{Sher} D. Sherrington, J. Phys. C: Solid State Phys. {\bf 12}, 5171 (1979);
Sherrington, in {\sl Les Houches, Session XXXI, Ill-Condensed
Matter} (North Holland Publishing Company, 1979); R. Johnston, D.
Sherrington, J. Phys. C: Solid State Phys. {\bf 15}, 3757 (1982)

\bibitem{SJS} S. John, H. Sompolinsky, M. Stephen,
Phys. Rev. B {\bf 27}, 5592 (1983)

\bibitem{ASZ} For review, see A. ALtland, B.D. Simons, and M.R. Zirnbauer,
Phys. Rep. {\bf 359}, 283 (2002).

\bibitem{Junits} After taking the continuum limit, $J$ now
becomes the spin-spin interaction constant times the square of the
lattice spacing.

\bibitem{GC} V. Gurarie, J. Chalker, Phys. Rev. Lett. {\bf 89},
136801 (2002)
\bibitem{Efetov} K. Efetov, {\sl Supersymmetry in Disorder and
Chaos}, Cambridge University Press



\end{thebibliography}
\end{document}